# Energy and environmental aspects of mobile communication systems

C. Lubritto[a], A. Petraglia[a,*], C. Vetromile[a], S. Curcuruto[b], M. Logorelli[b],  G. Marsico[b],  A. D'Onofrio[a].

[1]*Second University of Naples, Department of Environmental Science, Via Vivaldi, 43, I-81100  Caserta, Italy.*
[2]*Institute for the Environmental Research, Via Brancati, I-00100 Rome, Italy.*
*\* Corresponding author. Tel.:+39 0823 274629 . E-mail address: antonio.petraglia@unina2.it (A. Petraglia).*

## Abstract

The reduction of the energy consumptions of a Telecommunication Power System represents one of the critical factors of the telecommunication technologies, both to allow a sizeable saving of economic resources and to realize "sustainable" development actions. The consumption of about one hundred base stations for mobile phones were monitored for a total of over one thousand days, in order to study the energy consumption in relation to the environmental, electric and logistics parameters of the stations themselves. It was possible to survey, then, the role of the mobile communication systems in the general national energy framework and to plot the best areas of intervention for saving energy and improving the environmental impact, showing the role played by air conditioning and transmission equipments. Finally, new transmission algorithms and the use of renewable energy based techniques have been tested.

## Introduction

Mobile Telecommunication applications have seen, in the last years, a remarkable increase in the number of installations and consequently there has been a relevant growth of energy consumptions. This is due to an ever-growing interest about new and reliable services in mobility calls with an increase of the BTS operational hours and traffic management, in order to guarantee the quality of the service anywhere and any time.
In other words improving the energy efficiency of telecommunication networks is not just a necessary contribution towards the fight against global warming, but with the rapidly rising prices of energy, it is becoming also a financial opportunity [1-3].
From a technical and economical point of view, possible interventions in the field are:
   a) energetic auditing for radio-telecommunication stations in different operational contexts (urban and rural areas, different periods in the year, different working load, etc.);
   b) interventions of efficiency increase and energy saving such as reduction of transmission apparatus consumptions, optimization of air conditioning consumptions, efficiency in the temperature control system;
   c) evaluation and development of interventions and technical solutions based on the local production of a part of the consumed energy, through the use of photovoltaic cells on the infrastructures themselves; analyses of possible uses of other renewable sources (e.g. wind micro turbines) generating energy usable for  systems located in areas not reached by the electricity network;
   d) analyses of the social and environmental advantages from the introduction of technologies based on renewable sources;
   e) environmental monitoring of the sites where prototypal solutions have been installed, aimed to compare the conditions before and after the intervention.

The typical wireless network can be viewed as composed by three different sections:
   • the Mobile Switching Center (MSC): switching and  interfacing to fixed network;
   • the Base Transceiver Station (BTS): which is used as interface between network and mobile terminals;
   • the Mobile Terminals, normally limited to the hand-held devices.

The key elements are the BTSs because their contribution is the most relevant for the total energy consumption. Indeed, as the number of core network elements is low, the total energy consumption due to core network is low. Moreover, the power consumption of mobile terminals has been optimized in the years and is comparatively very low (a few Watts).
The energy allocation per function within the BTS has been extensively studied [1,4-5]. More than 60% of the power is consumed by the radio equipment and amplifiers, 11% is consumed by the DC power system and 25% by the cooling equipment, an air conditioning unit. The Radio Equipment and the Cooling are the two major sections where the highest energy savings potential resides. Other strategies, involving rearrangement of the network topology, are depicted in [6] and references therein.
Therefore it is very important to consider BTS Energy Savings Strategies applied both to the radio equipment, i.e. radio "standby" mode [7-9] and to the cooling, i.e. passive cooling, advanced climate control [10], and power equipment. Moreover one has to consider the aggregate effect that represents a further benefit [1].

Another way to reduce cost and $CO_2$ emissions is the evaluation and development of interventions and technical solutions based on the production of a part of the energy used by radio-telecommunication apparatuses, through the use of renewable sources (e.g. photovoltaic cells, wind micro turbines or new alternative power based on fuel cells) installed on the infrastructures themselves usable both by grid-connected and by off-grid telecommunication power systems. The use of alternative energy sources has been studied in particular for sites that are beyond the reach of an electricity grid, or where the electricity supply is unreliable or sites that are remote enough to make the regular maintenance and refueling of diesel generators prohibitive [11-12]. The choice of alternative energy sources will depend on local conditions, BTS typology and energy consumption amount. In most cases an hybrid solution combining solar and wind is actually the most feasible solution for autonomous BTSs site. Anyway, the size of solar cell and wind turbine have to be defined based on the BTS load and on-site availability of sun and wind as we will see later on.

In the following the results of on site experiments concerning energy consumption of BTSs, power saving actions and application of renewable energy supply for BTSs will be shown.

**Radio Base Station Energy consumptions.**

Energy auditing of a BTS is the most important step in the understanding of an energy management of wireless telecommunication power systems. With this aim we made a campaign of measurements for a radio-telecommunication apparatus starting from on-site measurements, performed in collaboration with Italian companies of mobile communications systems (Vodafone, H3G, Telecom and Wind).

Thanks to this collaboration, it has been possible to retrieve data coming from a statistical sample of 95 Radio Base Stations spread on the whole Italian national territory, that corresponds to more than 1000 monitoring days. All the field measurements were performed using a specific monitoring system described in chapter II of [13]: it is composed of a series of Remote Terminal Units (one for every station) collecting all the data of interest and sending them periodically to a Site Energy Saving SW tool installed in the Control Center Server.

Aim of our studies is to find statistical correlations between the energy consumption and the operational parameters of the BTS. Moreover we were interested to study both the energy consumption correlated to the transmission function of the apparatuses and the energy consumption related to the cooling of the equipments and infrastructures. To achieve this goal, we made extended statistical analysis (using the software "R" of the "R-Foundation for Statistical Computing" [14]).

We considered separately the following characteristics of the systems:
- Systems typologies (Shelter, Room, Outdoor);
- Systems technologies (GSM, DCS, UMTS);
- Localization (South, Centre and North Italy);

and the following operating parameters:
- Energy consumption (Wh);
- Instant Power (W);
- Internal temperature (°C);
- External temperature (°C);
- Phone traffic for cells ('erlang').

A specific database has been built. The results are:
- The average yearly consumption of a BTS is ca. 35500 kWh, considering that in Italy there are about 60.000 BTSs [15], the total average yearly consumption of the Italian BTS systems is ca. 2,1 TWh/year, which is the 0,6 % of the whole national electrical consumption (data source: TERNA 2007). In terms of economical and environmental impact, the data correspond to ca. 300M€ yearly energy costs and ca. 1,2 Mton of $CO_{2eq}$ emitted in the atmosphere every year.
- Carrying out analyses on the average energy consumptions associated to the different technologies, we note that the GSM energy consumptions are considerably higher than the UMTS technology - as it is expected because of the different mode of operation of the two technologies (Table 1).

**Energy Consumption/technology**

| Technology | kWh/day | kWh/year |
|---|---|---|
| UMTS | 72,97 | 26268 |
| GSM | 111,35 | 40085 |

*Table 1: Energy consumption for GSM and UMTS technology*

In Figure 1 we display the typical daily trend of the power consumptions of a BTS; we choose a day with large temperature excursion. We can clearly distinguish two different trends: a constant power value of about 3.5 kW for the night time and the morning, and an oscillating trend, with an average value of about 5 kW for the hottest hours and the late evening. This is a not a surprising trend if we consider that in the first case the consumption is due only to the transmission functions (constant trend), whereas in the second case besides the transmission functions there is the conditioning energy consumption, with a "saw tooth" behaviour generated by the on/off switching mechanism of the air-conditioning systems. The energy consumption can thus be divided in two parts: approx. 2/3 of the daily consumed energy is due to the transmission, whereas 1/3 of the energy is used to run the conditioning systems.

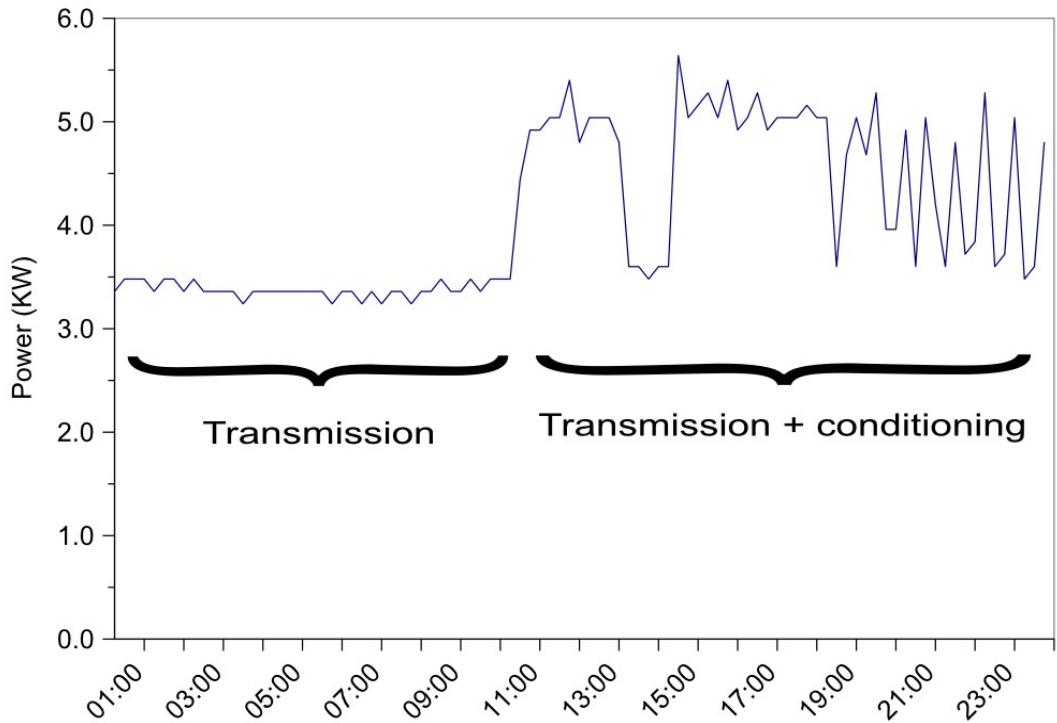

*Figure 1. Daily Energy Consumption of a BTS.*

Careful studies show a clear difference among the distribution functions on the coldest days, in which the instant power is presumably due only to the transmission function and the distribution functions on the warmest days, in which there are both the transmission and air conditioning contributions.

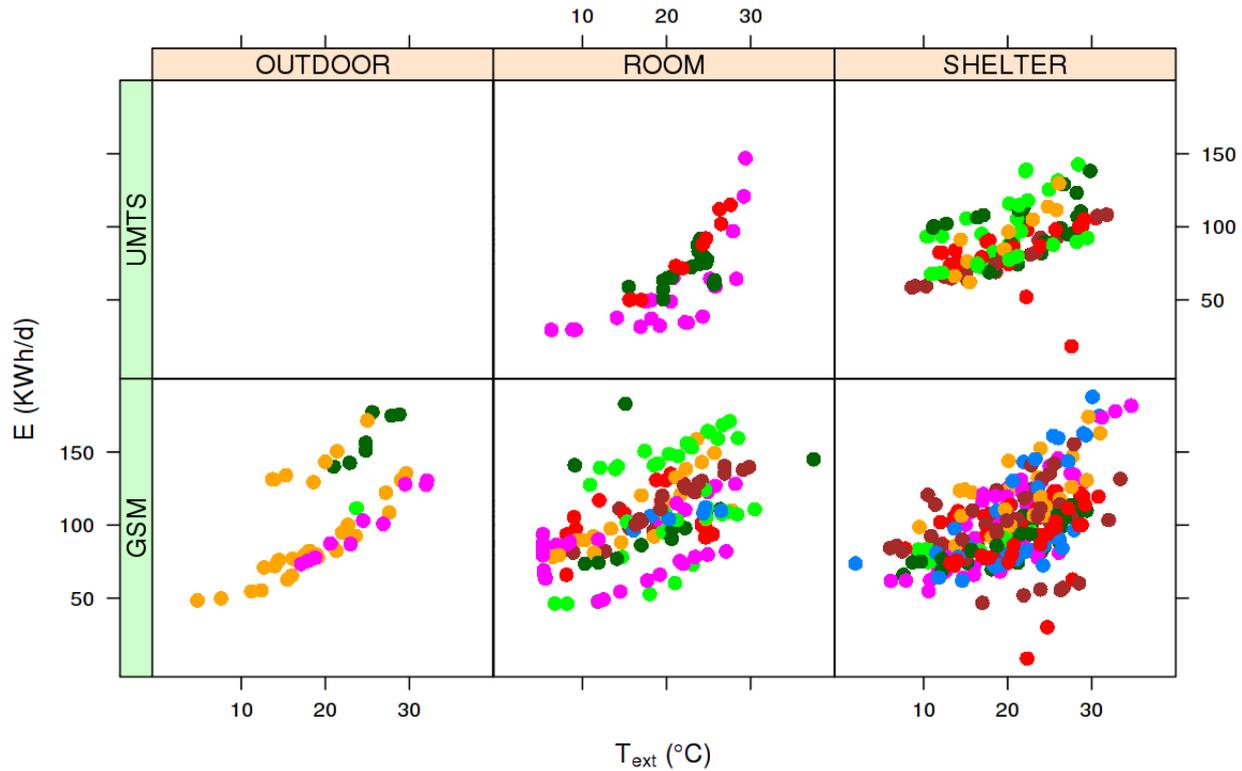

*Figure 2. Energy Consumption versus external temperature. Each color corresponds to a different BTS. There are no data for the UMTS/outdoor.*

In Figure 2 the variation of the energy consumptions versus the external temperature are plotted, for different BTS typology (shelter, room, outdoor) and BTS technology (UMTS and GSM), for 77 sites, for which there were reliable temperature information. An "universal trend" of the energy consumption versus the temperature can be recognized and it appears to be independent both from the technology used and from the typology of the BTS. Furthermore the higher dispersion of the consumption data in the room and shelter typologies, compared to the outdoor one, can be explained if we consider that in the first two cases there is a higher thermal dispersion – with a consequent increase of the consumption – which is needed to condition the equipments as well as the environment where they are installed.

Comparing the graphs for the room and shelter typology, it is clear that the UMTS technology has, on average, a lower energy consumption than GSM technology, because of the different characteristic of the mobile communication standards. No sizeable difference was found for energy consumptions for the three BTSs typologies.

In figure 3 we show the consumed energy per day for three Italian mobile operators versus the difference between external and internal shelter temperatures. It is clear that the superimposed regression lines have similar trend for the three operators. This is confirmed by the linear regression results shown in Table 2: both the intercepts and the slopes are compatible; indeed, p-values of the null-hypothesis of no difference between operator B and C and the operator A, taken as reference, are always greater than 0.15. Therefore we infer:

i) an energy consumption of about 110 kWh/day, when there is no difference between the external and internal shelter temperature;

ii) an increase of 2.1 kWh/day for each degree of temperature, which means that the maximum registered difference in temperature (25°C, as can be seen from the figure) gives an excursion in consumption of about 50 kWh/day.

| *Operator* | *Intercept (kWh/d)* | | *Slope (kWh/(d (Te-Ti)))* | | *p-value* | |
|---|---|---|---|---|---|---|
| | *Value* | *st. err.* | *value* | *st. err.* | *Intercept* | *slope* |
| A | 109,75 | 1,72 | 2,08 | 0,23 | | |
| B | 127,80 | 8,63 | 2,19 | 0,50 | 0.69 | 0.98 |
| C | 94,57 | 9,59 | 1,88 | 0,71 | 0.15 | 0.80 |

*Table 2. Linear regression summary of the results of figure 2.*

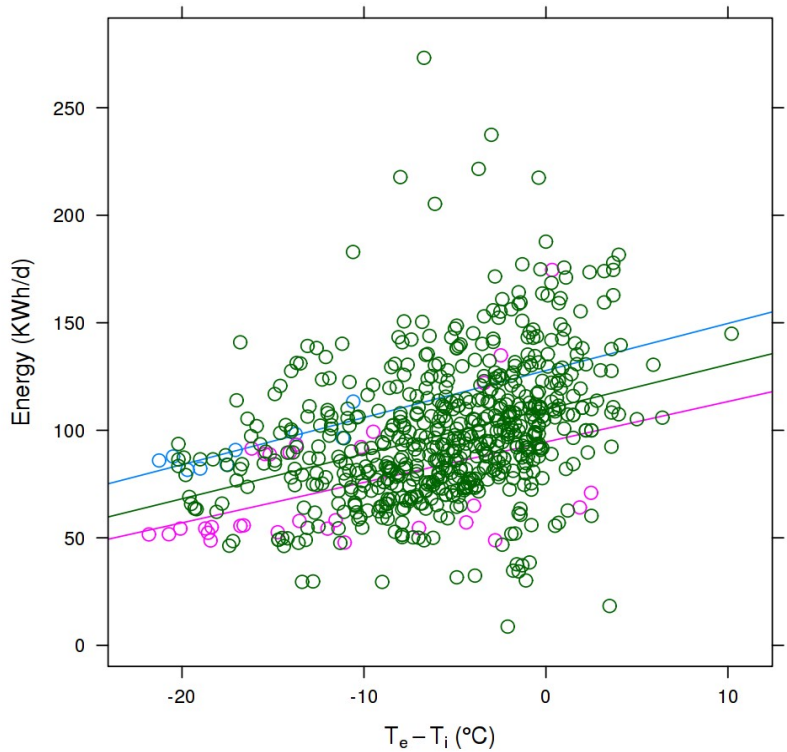

*Figure 3. Energy consumption versus temperature difference for the three Italian operators. A linear regression line is also shown. Each colour correspond to a different BTS operator.*

In order to understand a possible correlation between the energy consumptions data and the numbers of phone calls of a BTS, we show, in figure 4, the energy consumptions as a function of call traffic intensity (in erlang). It can be seen that there is no correlation between these two parameters; that means that at the moment no intervention which may regulate the BTS turning on/off when the call traffic goes down, has been realized.

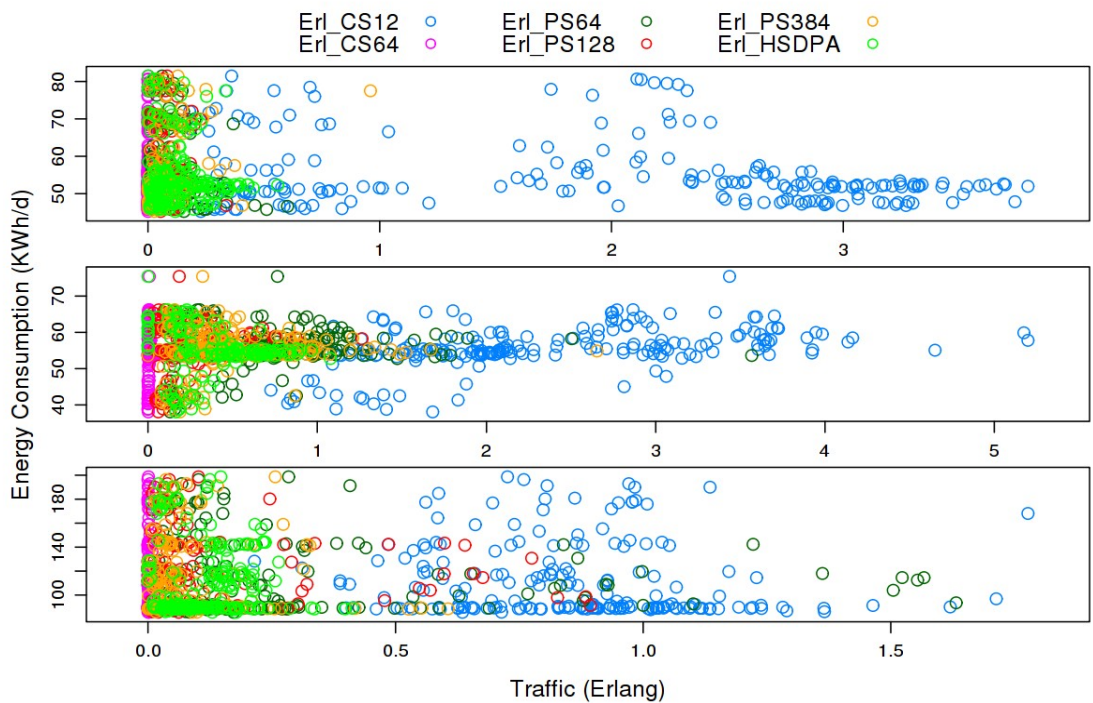

*Figure 4. Energy consumption vs traffic for three different sites. Different colours have been used for each transmission channel.*

# Energy saving: an experimental study

Starting from the statistical analyses on the BTSs energy consumptions, it is useful to study the possible interventions able to optimize and save energy consumptions. Our aim is to address useful actions both for saving on conditioning systems and on transmission consumption.

Regarding energy consumptions relevant to the air-conditioning we studied two possible intervention strategies. The first one was based on "intelligent" algorithms for the optimization and dynamical regulation of the air-conditioning functions. The second one was founded on the local cooling of the single electronic equipment.

Both hypotheses were based on the fact that inside the shelter the thermo-dynamic parameters (temperatures, humidity, etc.) can vary in ranges larger than in areas frequented by people. Therefore both the use of air-conditioning intelligent systems and the local cooling technologies are useful strategies for saving energy, exploiting the possibility to eliminate useless conditioning actions of the environment and mechanical parts. One can estimate that such interventions can achieve an energy saving from 5% to 10% of the air-conditioning consumptions, as it is confirmed by existing literature [1,16-18].

For optimizing the consumptions due to the transmission functions, we studied and tested a software feature launched by Ericsson that helps to use the BTS-GSM transmission power in a more efficient way [19]. These algorithms can correlate the phone traffic of a BTS and the energy consumptions. During periods of low network traffic, the feature effectively puts transceivers that are not being used in stand-by mode - overcoming the traditional practice of having radio equipment continually turned on, which results in an energy being waste.

It has been carried out an investigation on test BTSs in order to have direct hints about the feasibility of the project, practical problems and the estimation of realistic savings. To reach this goal a field measurements have been carried out in an operating BTS during periods in which the software feature was either activated or not activated.

Parameters useful in this kind of analysis are telephone traffic, BTS typology and location, number of transmitters and specific consumption parameters: energy consumed for each transmission channel and for each time-slot.

Measurements of energy consumptions and relevant environmental parameters have been realized in a suburban BTS located in Agliana (Toscana), composed of 3 GSM transmitters and 3 DCS transmitters. On this BTS the "power saving" function has been activated in field test of "power saving algorithms".

Results obtained by our monitoring actions are presented in Figure 5 for a DCS transmitter; a similar behaviour has been obtained for the other DCS and GSM transmitters. We report the energy consumption of the base radio station for three weeks; in two weeks the "Power Saving" function was activated (red and blue line) while in another was switched off (red line). It is clear that average values of energy consumptions are lower in the days in which the power saving function is active: a decrease of more than 10% of energy consumption is observed.

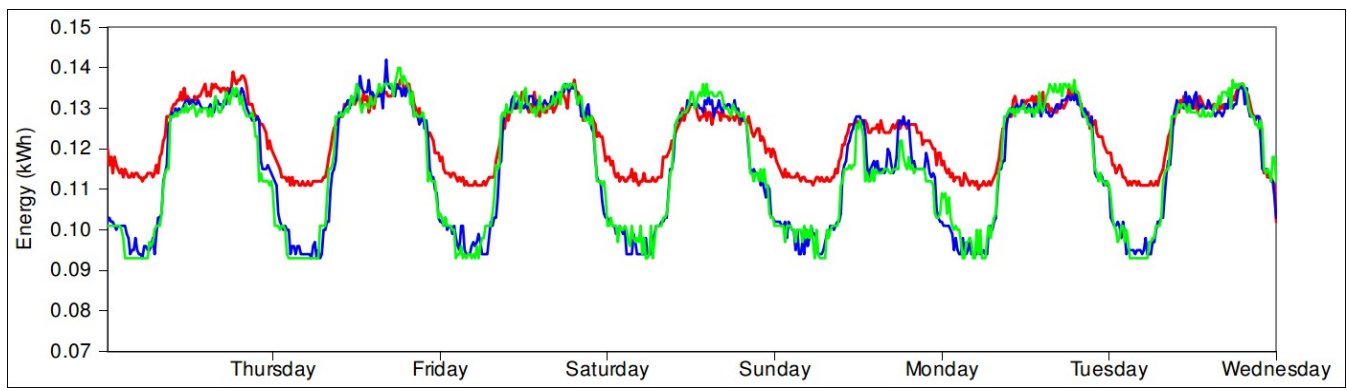

*Figure 5. Energy consumption of the Agliana BTS during the investigation time interval. Three different weeks are shown: power saving off (red) and power saving on with two slightly different sets of parameters (green and blue).*

The most of the energy saving is achieved during the night time - when the call traffic goes down and the "power saving" algorithm can switch off many transmitters. On the contrary, during the daytime, the curve trends coincide, since the high traffic does not allow the switch off of any transmitter. One can note the different behaviour in the weekends where the peaks in energy demand is lower, due to an average minor number of calls, and a small saving is possible also during daytime.

As it is not practically feasible to explore all the BTSs power consumption and to measure the power saved with different configurations of the feature algorithm, a simulation study will be carried out, in which the experimental data will be compared to the simulation analyses realized through a specific software which has been implemented and the best parameters ensuring good communications and best savings will be pointed out [20].

In conclusion, if we sum up the contributions of the energy saving of the air-conditioning and the transmission functions, – considering the feedback processes of the two components – we obtain a total yearly energy saving of 20% of the total consumptions (i.e. 7000 KWh per year) with an economical advantage of approximately 1000 €/year saved for each BTS, that means a further environmental advantage of non emitted 4 Tons of $CO_{2eq}$ in the atmosphere for each year and BTS.

# Telecommunications power systems and renewable energy

In order to introduce clean technology in the telecommunication power system management, one has to consider the use of renewable sources technologies (photovoltaic, wind, hybrid systems) installed on telecommunication system infrastructures. We analysed the most advanced technological solutions in the photovoltaic sector (single crystal, multi-crystal panels, amorphous silicon, thin films) and in the other renewable sources, useful to produce energy in situ according to the functioning conditions and the structural features of a radio-telecommunication station.

We chose to analyse the interventions in sites with different testing conditions for each provider, in order to plan and realize photovoltaic systems for basic radio stations in urban and rural areas (raw-land) and to analyse the use of different renewable sources for grid-off BTSs (not connected to the electrical net).

Two rural sites in which a photovoltaic plant has been built in collaboration with Vodafone provider, in order to understand how to use infrastructures of the BTS to obtain a total or partial architectural integration of the photovoltaic plants on the shelter or support pole. It has been shown that their energetic productivity depends on the geographical location, on the surface available to implement the photovoltaic plants and on the effects of shadow.

In two of these pilot sites, photovoltaic plants have been realized both on shelter and on the infrastructures; the area of PV modules varies from 16 to 20 $m^2$, limited by the available site space, to guarantee a yield of 2.0 and 2,5 kWp.

Figure 6 shows one of the installation with photovoltaic panels.

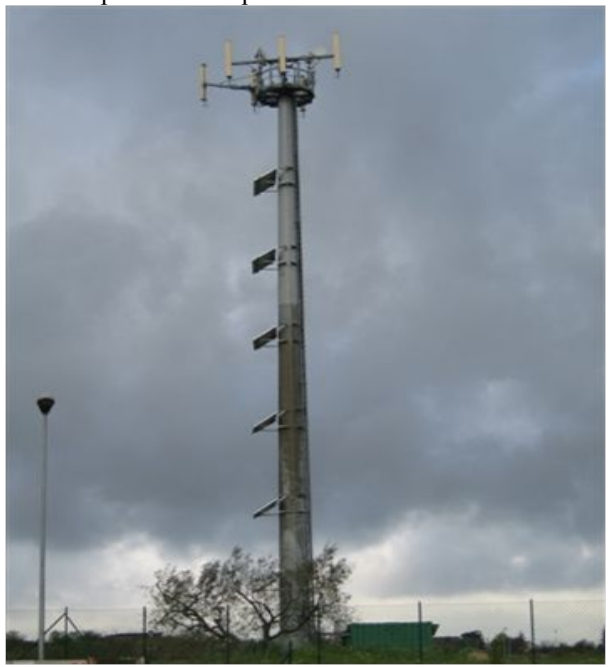

*Figure 6. A Photovoltaic installation over a BTS.*

These systems came into operation on 01/01/2008 and monitored until 30/05/2008, producing respectively 1100 and 1200 kWh; this implies an annual estimated production of 2640 and 2880 kWh. A further important element of the application - made by the involved provider - was to activate and complete the whole proceeding in order to obtain incentives by Italian electricity Authority (i.e. "Conto Energia" fund). It is to be noted that such an application gives an environmental advantage of approximately 3 Ton of not emitted $CO_{2eq}$/year for each BTS, besides the reduction of the pollution coming from further physical agents.

On these same sites we studied other technological solutions in the photovoltaic field. In particular very important is to evaluate the possibility to implement different photovoltaic technologies (amorphous, thin film) and to evaluate the technical feasibility and the investment advantage for each solution.

A very interesting application of renewable energy source with telecommunication systems is given by BTSs located far from the electrical network (grid-off systems). In order to analyse innovative technological solutions to produce energy close to the BTS, we studied technical characteristic of a BTS located on Gardaininu Mount (Nuoro, Sardinia), where an hybrid solution has been planned. It is composed by:

- a photovoltaic system with panels mounted on a support between the transmission tower and the south side of the external border, since we noticed the existence of an available surface of ca. 50 $m^2$.;
- a micro-wind system (power of 20kW), for which one can estimate energy production starting from wind average speed and hours of productivity in the specific site;

In the aforementioned site in Sardinia, the annual average speed of the wind is 5 m/s (CESI Wind Atlantis, 2007) and the annual productivity is 2000 hours. Consequently, the expected electric energy produced is 31567 kWh.

## Telecommunications power systems and environmental impacts

In the framework considered in the present paper, it becomes very important to study the relation between energetic aspects, environmental impacts and radio-telecommunication power systems.
One can consider at least three different relevant contexts:
- Impact of BTSs on the landscape;
- Electromagnetic pollution generated by the BTSs;
- Environmental impact coming from further polluting agents (emission of greenhouse gases, noise, etc.).

Regarding the impact on the environment, it is fundamental to build BTSs integrated in their territorial context. The possibility to use the structures of the plants (support poles, shelter, etc.) in order to realize energy production systems deriving from renewable sources could be an useful way to satisfy such request. It is evident that up to now such problems have not been faced, but we believe it is essential to define the right prescriptions in order to minimize the impact of the BTSs on the environment and, at the same time, to implement innovative power energy solutions.

Besides, the possibility to decrease the emission amount of the greenhouse gases deriving from the BTS activity is strictly joined to the introduction of the distributed production system of renewable energy. In fact, the use of solutions based on renewable energy, especially in grid-off conditions, can considerably reduce the impact on the environment caused by the station and the feeding generators management. The usage of a micro-wind or photovoltaic systems may avoid the implementing of feeding systems – which should be fed with fuel every day with the relevant high use of vehicles. That would mean a remarkable reduction of the emissions in the atmosphere.

Even more important is the impact on the environment and on the population of the electromagnetic fields generated by the transmission systems. Especially during the last years, such a problem has attracted the attention of single citizens and of the entire civil community, because of the considerable diffusion on the territory of low frequency electromagnetic fields sources (power lines) and of high frequency sources (radio-television stations and mobile phone stations). From a technical point of view, we should note that the field level in an area can considerably change according to the status of the existing transmitters, to the emitted power (at its turn variable according to the user needs) and to the further territorial characteristics, such as the presence of buildings and/or other obstacles which determine time-dependent reflections/refractions. Based on these premises and considering the assumed energy power saving strategies, we evaluated the feedback on the minimization of the electromagnetic fields emitted by the BTS. It is easy to understand that the use of a "power saving" system can give a valid contribution to the reduction of the BTS emissions. In fact, switching off the transmitters when the traffic goes down, means to get a null emission of the electromagnetic fields and a reduction of the daily average value emitted by the radio base stations. If we consider that, after the activation of the "power saving" algorithm there is a switch off of 70% of transmitters during night-time (from 24 to 8), we can estimate a daily average reduction of the electromagnetic emissions of 15-20%.

## Conclusions

In this paper results of on site experiments concerning energy consumption, power saving actions and application of renewable energies for BTSs supply have been studied. In particular a software system has been tuned and tested on selected BTSs showing the possibility of an energy saving of around 10-20 %. Distributed energy production from renewable sources (Photovoltaic, micro-wind mills) installed on or near the BTS has been considered and tested in selected case showing that a sizeable part of the needed electrical energy can be produced in this way avoiding $CO_2$ emission. Extensive application of the software system and of the renewable energy systems are foreseen.

## Acknowledgments

Results presented in this paper have been obtained in the framework of the research project "Telecommunication power systems: energy saving, renewable sources and environmental monitoring" realized by the Department of Environmental Sciences of the Second University of Naples (DSA-SUN) and the Advanced Institute for the Environmental Protection and Research (ISPRA), with the participation of the Italian suppliers of mobile telecommunications (H3G, Vodafone, Telecom and Wind) and their technological partners (Ericsson).
We thank Laura Miglio and Floriana Caterina for help and useful discussions.